\documentclass[11pt]{article}
\usepackage[pdftex]{graphicx,color}
\usepackage{fullpage,amssymb, amsmath, paralist}
\usepackage{epsfig,epstopdf}
\newtheorem{lemma}{Lemma}

\newtheorem{theorem}{Theorem}

\newenvironment{proof}{{\it Proof:\/}}{\hspace*{1em}\hfill$\Box$\vskip 0.1in}

\title{Nash equilibria in Voronoi games on graphs}
\author{Christoph \textsc{D\"urr}\thanks{LIX, CNRS UMR 7161, Ecole Polytechnique
        91128 Palaiseau, France.
         Supported by ANR Alpage.}
\and
        \textsc{Nguyen Kim} Thang\footnotemark[1]}

\begin{document}
\maketitle
\begin{abstract}
In this paper we study a game where every player is to choose a
vertex (facility) in a given undirected graph. All vertices
(customers) are then assigned to closest facilities and a player's
payoff is the number of customers assigned to it. We show that
deciding the existence of a Nash equilibrium for a given graph is
$\mathcal{NP}$-hard which to our knowledge is the first result of
this kind for a zero-sum game.  We also introduce a new measure,
the \emph{social cost discrepancy}, defined as the ratio of the
costs between the worst and the best Nash equilibria. We show that
the social cost discrepancy in our game is $\Omega(\sqrt{n/k})$
and $O(\sqrt{kn})$, where $n$ is the number of vertices and $k$
the number of players.
\end{abstract}

\section{Introduction}

Voronoi game is a widely studied game which plays on a continuous
space, typically a 2-dimensional rectangle. Players alternatively
place points in the space. Then the Voronoi diagram is
considered. Every player gains the total surface of the Voronoi cells
of his points~\cite{AhnCheng:Competitive-facility}.  This game is
related to the facility location problem, where the goal is to choose
a set of $k$ facilities in a bipartite graph, so to minimize the sum of
serving cost and facility opening cost~\cite{Vetta:facility}.

We consider the discrete version of the Voronoi game which plays
on a given graph instead on a continuous space. Whereas most
papers about these 
games~\cite{Cheong.Har-Peled:voronoi,FeketeMeijer:The-one-round-Voronoi} study the
existence of a winning strategy, or computing the best strategy
for a player, we study in this paper the  Nash equilibria.

Formally the discrete Voronoi game plays on a given undirected
graph $G(V,E)$ with $n = |V|$ and $k$ players. 
Every player
has to choose a vertex (\emph{facility}) from $V$, and every
vertex (\emph{customer}) is assigned to the closest facilities. A
player's payoff is the number of vertices assigned to his
facility. We define the \emph{social cost} as the sum of the
distances to the closest facility over all vertices.

We consider a few typical questions about Nash Equilibria:
\begin{itemize}
    \item Do Nash equilibria exist?
    \item What is the computational complexity for finding one?
    \item If they exist, can one be found from an arbitrary
        intital strategy profile with the best-response dynamic?
    \item It they exist, how different are their social costs?
\end{itemize}
The existence of Nash equilibria is a graph property for a fixed
number of players, and we give examples of graphs for which there
exist Nash equilibria and examples for which there are none. We
show that deciding this graph property is an $\mathcal{NP}$-hard
problem.

For particular graphs, namely the cycles, we characterize all Nash
equilibria. We show that the best-response dynamic does not
converge to a Nash equilibria on these graphs, but does for a
modified version of this game.

Finally we introduce a new measure. Assume that there are Nash
equilibria for $k$ players on graph $G$. Let $A$ be the largest
social cost of a Nash equilibrium, $B$ the smallest social cost of
a Nash equilibrium, and $C$ the smallest social cost of any
strategy profile, which is not necessarily an equilibrium.  Then
the ratio $A/C$ is called the price of anarchy
\cite{Koutsoupias.Papadimitriou:anarchy}, $B/C$ is called the
price of stability \cite{Anshelevich:stability}. We study a
different ratio $A/B$, which we call the \emph{social cost
discrepancy} of $G$. The social cost discrepancy of the game is
defined as the worst discrepancy over all instances of the game.
The idea is that a small social cost discrepancy guarantees that
the social costs of Nash equilibria do not differ too much, and measures a degree of choice in the game.  
In some settings it may be unfair to compare the cost of a Nash equilibrium with the optimal cost, which may not be attained by selfish agents.
Note that this ratio is upper-bounded by the price of anarchy. We
show that the social cost discrepancy in our game is
$\Omega(\sqrt{n/k})$ and $O(\sqrt{kn})$. Hence for a constant
number of players we have tight bounds.

\section{The game}

For this game we need to generalize the notion of vertex partition of a graph: A \emph{generalized partition} of a graph $G(V,E)$ is a set of $n$-dimensional non-negative vectors, which sum up to the vector with $1$ in every component, for $n=|V|$.

\begin{figure}[htb]
  \centerline{\input{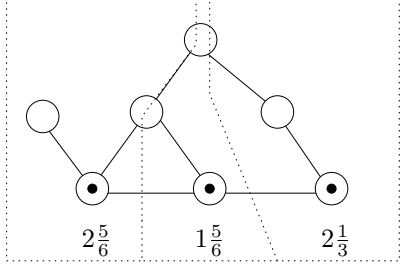}}
  \caption{A strategy profile of a graph (players are dots) and the corresponding payoffs.}
  \label{fig:gain}
\end{figure}

The Voronoi game on graphs consists of:
    \begin{itemize}
        \item
            A graph $G(V,E)$ and $k$ players. 
            We assume $k<n$ for $n=|V|$, otherwise the game has a trivial structure.
            The graph induces a distance between vertices $d:V\times V\rightarrow \mathbb N\cup \{\infty\}$,
            which is defined as the minimal number of edges of any connecting path, 
            or infinite if the vertices are disconnected.
        \item
            The strategy set of each player is $V$.
            A strategy profile of $k$ players is a vector $f=(f_{1},\ldots,f_{k})$
            associating each player to a vertex. 
        \item
            For every vertex $v\in V$ --- called \emph{customer} ---
            the distance to the closest facility is denoted as
            $d(v,f):=\min_{f_{i}} d(v,f_{i})$.
            Customers are assigned in equal fractions to the closest facilities as follows.
            The strategy profile $f$ defines the generalized partition $\{F_{1},\ldots, F_{k}\}$,
	    where
            for every player $1\le i\le k$ and every vertex $v\in V$,
            \[
            	F_{i,v} = \left\{ \begin{array}{cl}
					\frac{1}{|\text{arg min}_{j} d(v,f_{j})|} 
					&\text{if }  d(v,f_{i}) = d(v,f)
\\
					0 & \text{otherwise.}
				\end{array}
				\right.
		\]
	    We call $F_{i}$ the \emph{Voronoi cell} of player $i$.
            Now the payoff of player $i$
            is the (fractional) amount of customers
            assigned to it (see figure~\ref{fig:gain}), that is
            \(
               p_{i} := \sum_{v\in V} F_{i,v}.
            \)
    \end{itemize}

The \emph{best response} for player $i$ in the strategy profile
$f$ is a vertex $f'_{i}\in V$ maximizing the player $i$'s payoff
in the strategy profile $(f_{-i},f'_{i})$ which is a shorthand for
the profile which equals $f$, except that the strategy of player
$i$ is $f'_{i}$. If player $i$ can strictly improve his payoff by
choosing another strategy, we say that player $i$ is \emph{unhappy}
in $f$, otherwise he is \emph{happy}. The \emph{best response
dynamic} is the process of repeatedly choosing an arbitrary
unhappy player, and change it to an arbitrary best response. A
\emph{pure Nash equilibrium} is defined as a fixed point to the
best response dynamic,  or equivalently as a strategy profile
where all players are happy. In this paper we consider only pure
Nash equilibria, so we omit from now on the adjective ``pure''.

We defined players' payoffs in such a way, that there is a subtle
difference between the Voronoi game played on graphs and the
Voronoi game played on a continuous surface. Consider a situation
where a player $i$ moves to a location already occupied by a
single player $j$, then in the continuous case player $i$ gains
exactly a half of the previous payoff of player $j$ (since it is
now shared with $i$). However, in our setting (the discrete case),
player $i$ can sometimes gain more than a half of the previous
payoff of player $j$ (see figure~\ref{fig:cycle}).

Also note that the best responses for a player in our game are
computable in polynomial time, whereas for the Voronoi game
in continuous space, the problem seems
hard~\cite{Dehne.Klein.Seidel:voronoi}.

A simple observation leads to the following bound on the players
payoff.
\begin{lemma}                           \label{lem:gain}
  In a Nash equilibrium the payoff $p_{i}$ of every player $i$ is bounded by $n/2k < p_{i} < 2n/k$.
\end{lemma}
\begin{proof}
  If a player gains $p$ and some other player moves to the same
  location then both payoffs are at least $p/2$.
  Therefore the ratio between the largest and the smallest payoffs
  among all players can be at most $2$.
  If all players have the same payoff, it must be exactly $n/k$,
  since the payoffs sum up to $n$. Otherwise there is at least one
  player who gains strictly less than $n/k$, and another player who gains
  strictly more than $n/k$. This concludes the proof.
\end{proof}

\section{Example: the cycle graph}

Let $G(V,E)$ be the cycle on $n$ vertices with $V=\{v_i: i\in
\mathbb Z_n\}$ and $E=\{(v_i, v_{i+1}): i\in \mathbb Z_n\}$, where
addition is modulo $n$. The game plays on the undirected cycle,
but it will be convenient to fix an orientation. Let $u_0,\ldots,u_{\ell-1}$ be the  distinct facilities
chosen by $k$ players in a strategy profile $f$ with $\ell\le k$,
numbered according to the orientation of the cycle.  For every
$j\in\mathbb Z_{\ell}$, let $c_{j}\geq 1$ be the number of players
who choose the facility $u_{j}$ and let $d_j\geq 1$ be the length
of the directed path from $u_{j}$ to $u_{j+1}$ following the
orientation of $G$. Now the strategy profile is defined by these
$2\ell$ numbers, up to permutation of the players. We decompose
the distance into $d_{j}=1+2a_{j}+b_{j}$, for $0\leq b_{j}\leq 1$,
where $2a_{j} + b_{j}$ is the number of vertices between
facilities $u_{j}$ and $u_{j+1}$. So if $b_j=1$, then there is a
vertex in midway at equal distance from $u_j$ and $u_{j+1}$.

With these notations the payoff of player $i$ located on facility
$u_{j}$ is
\[
p_{i}:=
    \frac{b_{j-1}}{c_{j-1}+c_{j}}
    + \frac{a_{j-1}+1+a_{j}}{c_j}
    + \frac{b_{j}}{c_{j}+c_{j+1}}.
\]

All  Nash equilibria are explicitly characterized by the following
lemma. The intuition is that the cycle is divided by the players
into segments of different length, which roughly differ at most by
a factor $2$. The exact statement is more subtle because several
players can be located at a same facility and the payoff is
computed differently depending on the parity of the distances
between facilities.
\begin{lemma} \label{lem:cycle}
For a given strategy profile, let $\gamma$ be the minimal payoff
among all players, i.e: $\gamma := \min\{ p_{i} | 1 \leq i \leq
k\}$. Then this strategy profile is a Nash equilibrium if and only
if, for all $j\in\mathbb Z_{\ell}$:
\begin{enumerate}[(i)]
   \item $c_{j} \leq 2$
   \item $d_{j} \leq 2\gamma$
   \item If $c_{j} = 1$ and $d_{j-1}
           = d_{j} = 2\gamma$ then $c_{j-1} = c_{j+1} = 2$.
   \item If $c_{j-1} = 2, c_{j}=1, c_{j+1} = 1$ then $d_{j-1}$ is odd.
    \\
    If $c_{j-1} = 1, c_{j}=1, c_{j+1} = 2$ then $d_{j}$ is odd.
\end{enumerate}
\end{lemma}

\begin{lemma}
On the cycle graph, the best response dynamic does not converge.
\end{lemma}
\begin{proof}
  Figure~\ref{fig:cycle} shows an example of a graph, where the best
  response dynamic can iterate forever.
\begin{figure}[htb]
  \centerline{\input{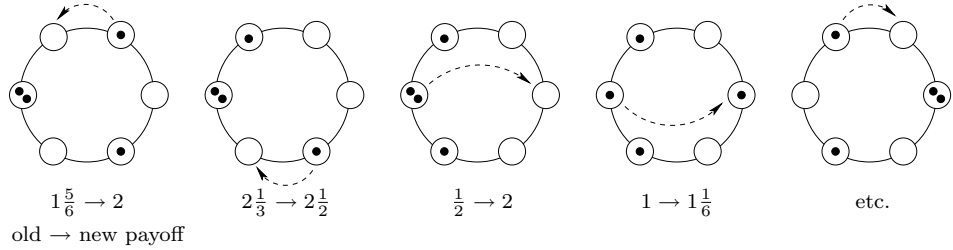}}
  \caption{The best response dynamic does not converge on this graph.}
  \label{fig:cycle}
\end{figure}
\end{proof}

However there is a slightly different Voronoi game in which the
best response dynamic converges~: The \emph{Voronoi game with
disjoint facilities} is identical with the previous game, except
that players who are located on the same facility now gain zero.

\begin{lemma}
On the cycle graph, for the \emph{Voronoi game with disjoint
facilities}, the best response dynamic does converge on a
strategy profile in which players are located on distinct
facilities.
\end{lemma}
\begin{proof}
 To show convergence we use a potential function. For this purpose we
 define the \emph{dominance order}: Let $A,B$ be two multisets.  If
 $|A|<|B|$ then $A\succ B$.  If $|A|=|B|\geq 1$, and $\max A > \max
 B$ then $A \succ B$.  If $|A|=|B|\geq 1, \max A = \max B$ and
 $A\backslash \{\max A\} \succ B\backslash\{\max B\}$ then $A\succ
 B$. This is a total order.

 The potential function is the multiset $\{d_{0}, d_{1}, \ldots,
 d_{k-1}\}$, that is all distances between successive occupied
 facilities.  Player $i$'s payoff --- renumbered conveniently ---
 is simply $(d_i+d_{i+1})/2$.  Now consider a best response for
 player $i$ moving to a vertex not
 yet chosen by another player, say between player $j$ and $j+1$.
 Therefore in the multiset $\{d_{0}, d_{1}, \ldots, d_{k-1}\}$, the
 values $d_i,d_{i+1},d_j$ are replaced by $d_i+d_{i+1},d',d''$ for
 some values $d',d''\geq 1$ such that $d_j = d'+d''$.  The new
 potential value is dominated by the previous one.  This proves that
 after a finite number of iterations, the best response dynamic converges to a Nash equilibrium.
\end{proof}


\section{Existence of a  Nash equilibrium is $\mathcal{NP}$-hard}

In this section we show that it is $\mathcal{NP}$-hard to decide
whether for a given graph $G(V,E)$ there is a  Nash
equilibrium for $k$ players. For this purpose we define a more
general but equivalent game, which simplifies the reduction.

In the \emph{generalized Voronoi game} $\langle G(V,E), U, w, k\rangle$  we are given a graph
$G$, a set of facilities $U\subseteq V$, a positive weight
function $w$ on vertices and a number of players $k$. Here the set
of strategies of each player is only $U$ instead of $V$. Also the
payoff of a player is the weighted sum of fractions of customers
assigned to it, i.e. the payoff of player $i$ is $p_{i}:=\sum_{v\in V} w(v) F_{i,v}$.

\begin{lemma}
  For every generalized Voronoi game $\langle G(V,E), U, w, k\rangle$ there is a
  standard Voronoi game $\langle G'(V',E'), k\rangle$ with $V\subseteq V'$,
  which has the same set of Nash equilibria and which is such that $|V'|$ is
  polynomial in $|V|$ and $\sum_{v\in V}w(v)$.
\end{lemma}
\begin{proof}
To construct $G'$ we will augment $G$ in two steps. Start with $V'
= V$.

First, for every vertex $u \in V$
such that $w(u)>1$, let $H_{u}$ be a set of $w(u)-1$ new vertices.
Set $V'=V' \cup H_{u}$ and connect $u$ with every vertex from $H_{u}$.

Second, let $H$ be a set of $k(a+1)$ new vertices where
$a=|V'|=\sum_{v\in V}w(v)$. Set $V'=V' \cup H$ and connect every
vertex of $U$ with every vertex of $H$.

Now in $G'(V',E')$ every player's payoff can be decomposed in the
part gained from $V'\backslash H$ and the part gained from $H$. We
claim that in a Nash equilibrium every player chooses a vertex
from $U$. If there is at least one player located in $U$, then the
gain from $H$ of any other player is $0$ if located in
$V'\backslash(U\cup H)$, is $1$ if located in $H$ and is at least
$a+1$ if located in $U$. Since the total payoff from $V'\backslash
H$ over all players is $a$, this forces all players to be located
in $U$.

Clearly by construction, for any strategy profile $f\in U^{k}$,
the payoffs are the same for the generalized Voronoi game in $G$
as for the standard Voronoi game in $G'$. Therefore we have
equivalence of the set of Nash equilibria in both games.
\end{proof}

\begin{figure}[htbp]
   \centerline{      \input{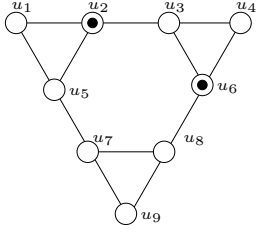}   }
   \caption{Example of a graph with no Nash equilibrium for 2 players.}
   \label{fig:triangle}
\end{figure}

Our $\mathcal{NP}$-hardness proof will need the following gadget.
\begin{lemma}                                   \label{lemma:triangle}
For the graph $G$ shown in figure \ref{fig:triangle} and $k=2$ players, there is no
 Nash equilibrium.
\end{lemma}
\begin{proof}
We will simply show that given an arbitrary location of one
player, the other player can move to a location where he gains at
least $5$. Since the total payoff over both players is $9$, this
will prove that there is no Nash equilibrium, since the best response
dynamic does not converge.

By symmetry without loss of generality the first player is located
at the vertices $u_{1}$ or $u_{2}$. Now if the second player is
located at $u_{6}$, his payoff is at least $5$.
\end{proof}

\begin{theorem}
Given a graph $G(V,E)$ and a set of $k$ players, deciding the
existence of Nash equilibrium for $k$ players on $G$ is
$\mathcal{NP}$-complete.
\end{theorem}
\begin{proof}
The problem is clearly in $\mathcal{NP}$, since best responses can
be computed in polynomial time, therefore it can be verified
efficiently if a strategy profile is a Nash equilibrium.

The proof of $\mathcal{NP}$-hardness is by the reduction from
\textsc{3-Partition}, which  is unary $\mathcal{NP}$-complete
\cite{GareyJohnson:Complexity-results}. In this later problem we
are given integers $a_{1},\ldots,a_{3m}$ and $B$ such that $B/4 <
a_{i} <B/2$ for every $1\leq i\leq 3m$, $\sum_{i=1}^{3m} = mB$ and
have to partition them into disjoint sets
$P_{1},\ldots,P_{m}\subseteq\{1,\ldots,3m\}$ such that for every
$1\leq j \leq m$ we have $\sum_{i\in P_{j}} a_{i} = B$.

We construct a weighted graph $G(V,E)$ with the weight function
$w: V \rightarrow \mathbb{N}$ and a set $U\subseteq V$ such that
for $k=m+1$ players ($m \geq 2$) there is a  Nash equilibrium to
the generalized Voronoi game $\langle G,U,w,k \rangle$ if and only
if there is a solution to the \textsc{3-Partition} instance. We
define the constants $c={3m \choose 3}+1$ and $d = \left\lfloor \frac{Bc
- c + c/m}{5} \right\rfloor + 1$. The graph $G$ consists of 3 parts. In
the first part $V_1$, there is for every $1\leq i\leq 3m$ a vertex
$v_{i}$ of weight $a_{i}c$. There is also an additional vertex
$v_0$ of weight $1$. In the second part $V_2$, there is for every
triplet $(i,j,k)$ with $1\leq i<j<k\leq 3m$ a vertex $u_{ijk}$ of
unit weight. --- Ideally we would like to give it weight zero,
but there seems to be no simple generalization of the game which
allows zero weights, while preserving the set of  Nash
equilibria. ---  Every vertex $u_{ijk}$ is connected to $v_0, v_{i},
v_{j}$ and $v_{k}$.  The third part $V_3$, consists of the 9
vertex graph of figure~\ref{fig:triangle} where each of the
vertices $u_{1},\ldots, u_{9}$ has weight $d$. To complete our
construction, we define the facility set $U:=V_2 \cup V_3$.

\begin{figure}[htbp]
   \centerline{      \input{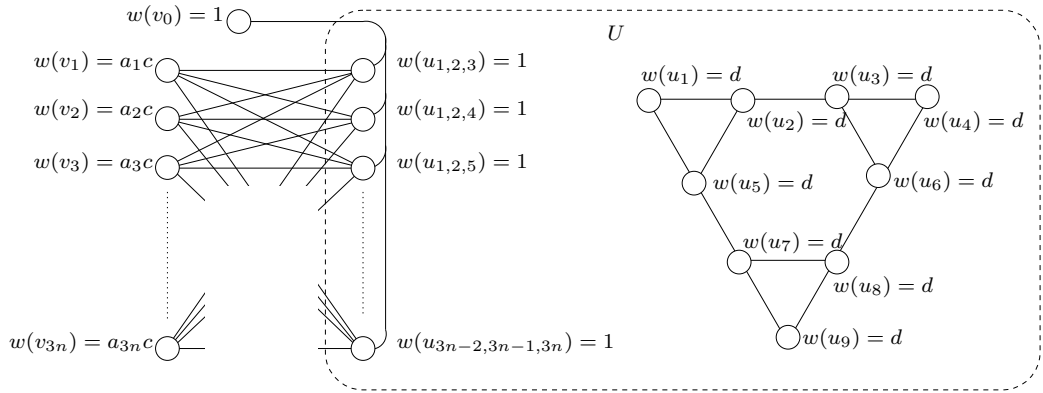}   }
   \caption{Reduction from \textsc{3-Partition}.}
   \label{fig:graph}
\end{figure}

First we show that if there is a solution $P_{1},\ldots,P_{m}$ to
the \textsc{3-Partition} instance then there is a  Nash
equilibrium for this graph.  Simply for every $1\leq q\leq m$ if
$P_{q}=\{i,j,k\}$ then player $q$ is assigned to the vertex
$u_{ijk}$. Player $m+1$ is assigned to $u_{2}$.  Now player
$(m+1)$'s payoff is $9d$, and the payoff of each other player $q$
is $Bc+c/m$.  To show that this is a  Nash equilibrium we need to
show that no player can increase his payoff.  There are different
cases. If player $m+1$ moves to a vertex $u_{ijk}$, his payoff
will be at most $\frac34Bc+c/(m+1)<9d$, no matter if that vertex
was already chosen by another player or not. If player $1\leq
q\leq 3m$ moves from vertex $u_{ijk}$ to a vertex $u_{\ell}$ then
his gain can be at most $5d<Bc+c/m$.  But what can be his gain, if
he moves to another vertex $u_{i'j'k'}$?  In case where
$i=i',j=j', k\neq k'$, $a_{i}c+a_{j}c$ is smaller than $\frac 34
Bc$ because $a_{i}+a_{j}+a_{k}=B$ and $a_{k}>B/4$. Since
$a_{k'}<B/2$, and player $q$ gains only half of it, his payoff is
at most $a_{i}c + a_{j}c + a_{k'}c/2 + c/m < Bc + c/m$ so he again
cannot improve his payoff. The other cases are similar.

Now we show that if there is a  Nash equilibrium, then it
corresponds to a solution of the \textsc{3-Partition} instance. So
let there be a  Nash equilibrium. First we claim that there is
exactly one player in $V_3$. Clearly if there are 2 players, this
contradicts equilibrium by Lemma~\ref{lemma:triangle}.  If there
are 3 players or more, then by a counting argument there are
vertices $v_i,v_j,v_k$ which are at distance more than one from
any player.  One of the players located at $V_3$ gains at most
$3d$ and if he moves to $u_{ijk}$, his payoff would be at least
$\frac 34Bc + c/m > 3d$.  Now if there is no player in $V_3$, then
any player moving to $u_2$ will gain $9d> \frac32 Bc+c/m$ which is
an upper bound for the payoff of players located in $V_2$.  So we
know that there is a single player in $V_3$ and the $m$ players in
$V_2$ must form a partition, since otherwise there is a vertex
$v_i \in V_1$ at distance at least 2 to any player. So, by the
previous argument, there would be a player in $V_2$ who can
increase his payoff by moving to the other vertex in $V_2$ as
well. (He moves in such a way that his new facility is at distance
1 to $v_i$.) Moreover, in this partition, each player gains
exactly $Bc + c/m$ because if one gains less, given all weights in
$V_1$ are multiple of $c$, he gains at most $Bc - c + c/m$ and he
can always augment his payoff by moving to $V_{3}$ ($5d > Bc - c +
c/m$).
\end{proof}

\section{Social cost discrepancy}

In this section, we study how much the social cost of  Nash
equilibria can differ for a given graph, assuming Nash equilibria
exist. We define the \emph{social cost} of a strategy profile $f$ as $\text{cost}(f):= \sum_{v\in V} d(v, f)$.  
Since we assumed $k<n$ the cost is always non-zero.
The \emph{social cost discrepancy} of the game is the maximal fraction $\text{cost}(f)/\text{cost}(f')$ over all Nash equilibria $f,f'$. 
For unconnected graphs, the social cost can be infinite, and so can be the social cost discrepancy. Therefore in this section we consider only connected graphs.

\begin{lemma}                               \label{lem:discrepancy}
There are connected graphs for which the social cost discrepancy is
$\Omega(\sqrt{n/k})$, where $n$ is the number of vertices and $k$ the number of players.
\end{lemma}
\begin{proof}
We construct a graph $G$ as shown in figure~\ref{fig:cost}. 
The total number of vertices in the graph is $n = k(2a+b+2)$.
We distinguish two strategy profiles $f$ and $f'$:  the vertices occupied by $f$ are marked with round dots, and the vertices of $f'$ are marked with triangular dots. 
\begin{figure}[htbp]
   \centerline{      \input{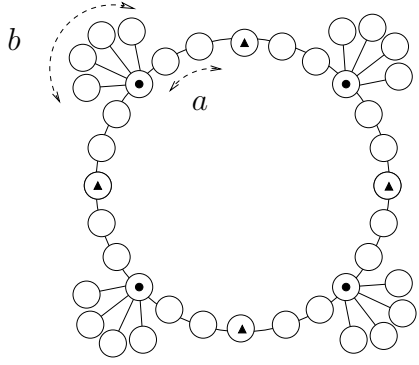}   }
   \caption{Example of a graph with high social cost discrepancy.}
   \label{fig:cost}
\end{figure}

By straightforward verification, it can be checked that both $f$ and $f'$
 are  Nash equilibria.  However the social
cost of $f$ is $\Theta(kb+ka^2)$ while the social cost of $f'$
is $\Theta(kab+ka^2)$.  The ratio between both costs is
$\Theta(a)=\Theta(\sqrt{n/k})$ when $b=a^2$ and thus the cost
discrepancy is lower bounded by this quantity.
\end{proof}

The \emph{radius} of the Voronoi cell of player $i$ is defined as
$\max_{v} d(v,f_{i})$ where the maximum is taken over all vertices
$v$ such that $F_{i,v} > 0$. The \emph{Delaunay
triangulation} is a graph $H_{f}$ on the $k$ players. 
There is an edge $(i,j)$ in $H_{f}$
either if there is a vertex $v$ in $G$ with $F_{i,v}>0$ and $F_{j,v}>0$ or if there is an edge $(v,v')$ in $G$  with $F_{i,v}>0$ and $F_{j,v'}>0$.

We will need to partition the Delaunay triangulation into small
sets, which are $1$-dominated and contain more than one vertex.
We call these sets \emph{stars}:
For a given graph $G(V,E)$ a vertex set
$A\subseteq V$ is a \emph{star} if $|A|\geq 2$, and there is a
\emph{center} vertex $v_0\in A$ such that for every $v\in A, v\neq v_0$ we have
$(v_0,v)\in E$.  Note that our definition allows the existence of
additional edges between vertices from $A$.

\begin{lemma}
For any connected graph $G(V,E)$, $V$ can be partitioned into
stars.
\end{lemma}
\begin{proof}
  We define an algorithm to partition $V$ into stars.

  As long as the graph contains edges, we do the following. We start choosing
  an edge: If there is a vertex $u$ with a unique neighbor $v$, then
  we choose the edge $(u,v)$; otherwise we choose an arbitrary edge
  $(u,v)$.  Consider the vertex set consisting of $u,v$ as well as of
  any vertex $w$ that would be isolated when removing edge $(u,v)$.  Add
  this set to the partition, remove it as well as
  adjacent edges from $G$ and continue.

  Clearly the set produced in every iteration is a star.  Also when
  removing this set from $G$, the resulting graph does not contain an
  isolated vertex. This property is an invariant of this algorithm,
  and proves that it ends with a partition of $G$ into stars.
\end{proof}

Note that, when a graph is partitioned into stars, the centers of
these stars form a
dominating set of this graph. Nevertheless, vertices in a
dominating set are not necessarily centers of any star-partition
of a given graph.

\begin{figure}[htbp]
   \centerline{      \input{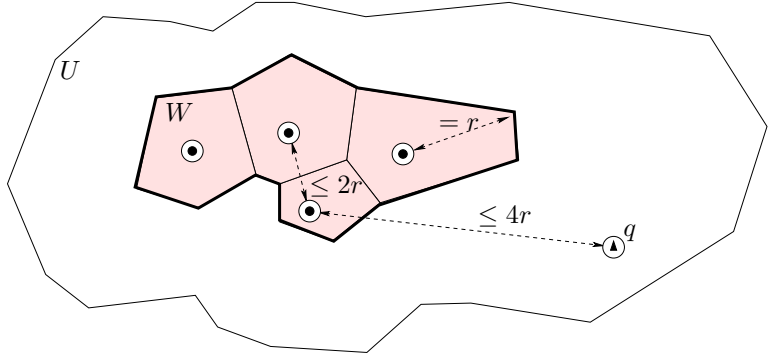}   }
   \caption{Illustration of lemma~\ref{lem:close}.}
   \label{fig:star}
\end{figure}

The following lemma states that given two different Nash
equilibria $f$ and $f'$, every player in $f$ is not too far from
some player in $f'$.  For this purpose we partition the Delaunay triangulation $H_{f}$ into stars, and bound the distance from any player of a star to $f'$ by 
some value depending on the star.

\begin{lemma}                               \label{lem:close}
Let $f$ be an equilibrium and $A$ be a
star of a star partition of the Delaunay triangulation $H_{f}$. 
Let $r$ be the maximal radius of the Voronoi
cells over all players $i\in A$.
 Then, for any
equilibrium $f'$, there exists a player $f'_{j}$ such that $d(f_{i},f'_j) \leq 6r$ for every $i \in A$.
\end{lemma}
\begin{proof}
Let $U =\{ v\in V: \min_{i\in A}d(v,f_{i}) \leq 4r\}$. If we can show
that there is a facility $f'_{j}\in U$ we would be done, since
by definition of $U$ there would be a player $i\in A$ such that
$d(f_{i},f'_{j})\leq 4r$ and the distance between any pair of facilities of
$A$ is at most $2r$. This would conclude the lemma.

So for a proof by contradiction, assume that in the strategy profile $f'$
there is no player located in $U$. 
Now consider the player with smallest payoff in $f'$.
His payoff is not more than $n/k$. However
if this player would choose as a facility the center of the star
$A$, then he would gain strictly more: By the choice of $r$, any
vertex in $W$ is at distance at most $3r$ to the center of the
star. However, by assumption and definition of $U$, any other
facility of $f'$ would be at distance strictly greater than $3r$
to any vertex in $W$. So the player would gain at least all
vertices around it at distance at most $3r$, which includes $W$.
Since any player's payoff is strictly more than $n/2k$ by
Lemma~\ref{lem:gain}, and since a star contains at least two
facilities by definition, this player would gain strictly more
than $n/k$, contradicting that $f'$ is an equilibrium. This
concludes the proof.
\end{proof}

\begin{theorem}                            \label{thm:discrepancy}
For any connected graph $G(V,E)$ and any number of players $k$ the
social cost discrepancy is $O(\sqrt{kn})$, where $n=|V|$.
\end{theorem}
\begin{proof}
Let $f,f'$ be arbitrary equilibria on $G(V,E)$.
We will consider a generalized partition of $V$ and for each part
bound the cost of $f'$ by $c\sqrt{kn}$ times the cost of $f$ for some constant $c$.

For a non-negative $n$-dimensional vector $W$ we 
define the cost restricted to $W$ as
$\textrm{cost}_{W}(f) =
\sum_{v\in V} W_{v}\cdot d(v,f)$. 
Now the cost of $f$ would write
as the sum of $\textrm{cost}_{W}(f)$ over the vectors $W$ from some
fixed generalized partition. 

Fix a star partition of the Delaunay
triangulation $H_{f}$. Let $A$ be an
arbitrary star from this partition, $a=|A|$, and $W$ be the sum of the corresponding Voronoi cells,
i.e. $W=\sum_{i\in A} F_{i}$.
We will show that $\textrm{cost}_W (f') = O(\sqrt{kn}\cdot \textrm{cost}_W (f))$, which would conclude the proof.
There will be two cases $k\le n/4$ and $k>n/4$.

By the previous lemma there is a vertex $f'_{j}$ such that
$d(f_{i},f'_{j})\leq 6r$ for all $i\in A$, where $r$ is the largest radius
of all Voronoi cells corresponding to the star $A$. So the cost of
$f'$ restricted to the vector $W$ is
\begin{align}\label{eq:costIneq}
    \textrm{cost}_W(f') &= \sum_{v \in V} W_{v}\cdot d(v,f')
    \leq \sum_{v \in V} W_{v}\cdot d(v,f'_{j})  \notag \\
    &= \sum_{v \in V} \sum_{i\in A} F_{i,v} \cdot d(v,f'_{j}) \notag \\
    &\leq \sum_{v \in V} \sum_{i\in A} F_{i,v} \cdot
        \left ( d(v,f_{i}) + d(f_{i},f'_{j})  \right ) \notag \\
    &\leq \textrm{cost}_{W}(f) + 6r\cdot|W|,
\end{align}
where $|W| := \sum_{v \in V} W_{v}$.

Moreover by definition of the radius, there is a vertex $v$ with
$W_{v} > 0$ such that the shortest path to the closest facility in
$A$ has length $r$.  So the cost of $f$ restricted to $W$ is
bigger than the cost restricted to this shortest path:
\[
    \textrm{cost}_W(f) \geq (\frac{1}{k}\cdot 1 + \frac{1}{k}\cdot 2 + \ldots
        + \frac{1}{k} \cdot r) \geq \frac{1}{k}\cdot r(r-1)/2.
\]
(The fraction $\frac{1}{k}$ appears because a vertex can be
assigned to at most $k$ players.)

First we consider the case $k \leq n/4$.  We have
\[
    \textrm{cost}_W(f) \geq |W| - |A| \geq a(n/2k - 1) \geq an/4k.
\]
The first inequality is because the distance of all customers
which are not facilites to a facility is at least one. The
second inequality is due to Lemma~\ref{fig:gain} and $|W|$ is
the sum of payoffs of all players in $A$.

Note that $|W| \leq n$ and $2 \leq a \leq k$ . Now if $r\leq
\sqrt{an}$, then
\[
    \frac{\textrm{cost}_{W}(f')}{\textrm{cost}_{W}(f)}
    \leq 1 + \frac{6r|W|}{\textrm{cost}_{W}(f)}
    \leq 1 + \frac{6r\cdot a\cdot2n/k}{an/4k}
    = O(r)=O(\sqrt{kn}).
\]
And if $r\geq \sqrt{an}$, then
\[
    \frac{\textrm{cost}_{W}(f')}{\textrm{cost}_{W}(f)}
\leq
    1 + \frac{6r|W|}{\textrm{cost}_{W}(f)})
    \leq 1 + \frac{6r\cdot a\cdot 2n/k}{r(r-1)/2k}
    = O(an/r)=O(\sqrt{kn}).
\]

Now we consider case $k>n/4$.  In any equilibrium, the maximum payoff
is at most $2n/k$. Moreover the radius $r$ of any Voronoi cell is upper bounded by $n/k+1$, otherwise the player with minimum gain (which is at most $n/k$) could increase his gain by moving to a vertex which is at distance at least $r$ from every other facility.
Therefore $r=O(1)$.  Summing~\eqref{eq:costIneq} over all stars with associated 
partition $W$,
we obtain $\textrm{cost}(f') \leq \textrm{cost}(f) + cn$, for some constant $c$.
Remark that the social cost of any equilibrium is at least $n-k$.
Hence, 
$\frac{\textrm{cost}(f')}{\textrm{cost}(f)} = O(n)$.
\end{proof}

%


\bibliographystyle{plain}
\bibliography{gametheory}

\vspace{2cm}
\section*{Appendix}
\setcounter{lemma}{1}
\begin{lemma} 
For a given strategy profile, let $\gamma$ be the minimal payoff
among all players, i.e: $\gamma := \min\{ p_{i} | 1 \leq i \leq
k\}$. Then this strategy profile is a Nash equilibrium if and only
if, for all $j\in\mathbb Z_{\ell}$:
\begin{enumerate}[(i)]
   \item $c_{j} \leq 2$
   \item $d_{j} \leq 2\gamma$
   \item If $c_{j} = 1$ and $d_{j-1}
           = d_{j} = 2\gamma$ then $c_{j-1} = c_{j+1} = 2$.
   \item If $c_{j-1} = 2, c_{j}=1, c_{j+1} = 1$ then $d_{j-1}$ is odd.
    \\
    If $c_{j-1} = 1, c_{j}=1, c_{j+1} = 2$ then $d_{j}$ is odd.
\end{enumerate}
\end{lemma}
\begin{proof}
\emph{(Necessary)} We will show that if a strategy profile does
not satisfy one of the conditions then it is not a Nash
equilibrium.
\begin{enumerate}[(i)]
   \item Suppose that there is a vertex $u_{j}$ with $c_{j} \geq 3$.
       Assume $d_{j-1}\leq d_{j}$, the other case is symmetric.
       Since there are vertices on the cycle which are not occupied by a player,
       the payoff of each player located on $u_{j}$ must be at least $1$.
       Therefore $d_{j}\geq 3$. Let $u'$ be the vertex immediately after $u_{j}$ in the cycle.
       We show now that one of the players located at $u_{j}$ can move to $u'$
       and strictly increase his payoff, which would contradict that the strategy
       profile is a Nash equilibrium.
       We decompose the distances into $d_{j-1} = 2a_{j-1} + b_{j-1} +
       1$ and $d_{j} = 2a_{j} + b_{j} + 1$ where $0 \leq b_{j-1}, b_{j}
       \leq 1$. By $d_{j}\geq 3$ we have $a_{j}\geq 1$.
       By $d_{j-1} \leq d_{j}$ we have $a_{j-1} \leq a_{j}$.
       Now the payoff of a player in facility $u_j$ is
       \[
       \frac{b_{j-1}}{c_{j-1} + c_{j}} +
        \frac{a_{j-1} + 1 + a_{j}}{c_{j}} +
        \frac{b_{j}}{c_{j} +
       c_{j+1}}.
    \]
    And if he moves to $u'$, his payoff would be
    \[
    a_{j}+b_{j} + \frac{1-b_{j}} {c_{j}}.
    \]
    In both cases $b_{j}=0$ or $b_{j}=1$ this new payoff is strictly greater.

   \item Suppose that there exists $j$ such that $d_{j} \geq 2\gamma
       + 1$. As previous, we decompose $d_{j} = 2a + b +1$ where $0 \leq b
       \leq 1$. Assume that $c_{j} \leq c_{j+1}$, the other case is symmetric.
       Let $u'$ be a vertex between $u_{j}$ and $u_{j+1}$ and at even distance
       to $u_{j+1}$.  Note that the number of vertices between $u_j$ and $u_{j+1}$
       except $u'$ is $2a + b -1 = 2(a+b-1) + (1-b)$. If a player moves to $u'$
       he will gain at least $a + b + \frac{1 - b}{1+c_{j}}$.
       Since $2a + b + 1 \geq 2\gamma + 1$,
       this payoff always greater than $\gamma$, so the player whose payoff is $\gamma$
       has an incentive to dislocate, contradicting that the strategy profile is a Nash
       equilibrium.

   \item Assume that $c_{j} = 1$, $d_{j-1} = d_{j} = 2\gamma$ and at
       least one of $c_{j-1}, c_{j}$ is 1.
       If a player whose payoff is $\gamma$ moves to facility $u_{j}$, he will gain
       at least $\gamma - \frac{1}{2} + \frac{1}{c_{j-1} + 2} + \frac{1}{c_{j} +
       2}$, which is strictly larger than $\gamma$.

   \item Assume $c_{j} = 1$ and $c_{j-1} = 2, c_{j+1} = 1$, the other case is symmetric.
       Now if  the distance $d_{j-1}$ is even, the player located at $u_{j}$ can stricly
       increase his payoff by moving to a vertex between $u_{j-1}$ and $u_{j}$
       and at odd distance to $u_{j-1}$. The idea is that if this player gets a
       fractional payoff from some mid-way vertex, he'll better off sharing it
       with one, rather than two players.
       \end{enumerate}

\emph{(Sufficient)} We will prove that if a strategy profile satisfies
all conditions, then it is a Nash equilibrium.
\begin{enumerate}[(i)]
   \item By (iii), if a player is alone on his facility then he has no
       incentive to make a \emph{local move} --- a move such that the set
       of his neighbors doesn't change.
   \item Without loss of generality, assume that $c_{j} \leq c_{j+1}$.
       If a player moves to a vertex between facilities $u_j$ and
       $u_{j+1}$, similarly as above, his payoff will be at most
       $1 + (a + b - 1) + \frac{1-b}{c_{j} + 1}$ where
       $d_{j} = 2a + b + 1$. Since $d_{j} \leq 2\gamma$, this new
       payoff will be at most $\gamma$ so the player has no incentive to move.
   \item If a player moves from facility $u_{i}$ to facility
       $u_{j}$ which is not his neighbor (i.e, $i \notin \{j-1,j\}$),
       he will gains $\frac{b_{j-1}}{c_{j-1} + c_{j} + 1} +
       \frac{a_{j-1} + 1 + a_{j}}{c_{j} + 1} +
       \frac{b_{j}}{c_{j} + 1 + c_{j+1}}$ where $d_{j-1} = 2a_{j-1} +
       b_{j-1} + 1$ and $d_{j} = 2a_{j} + b_{j} + 1$. Again, since
       (iii), this payoff is less or equal $\gamma$.
   \item Suppose that a player $j$ (from facility $u_{j}$) moves to
       one of his facility-neighbor $u_{j+1}$.
       If $c_{j} > 1$ then the situation is as in the above
       paragraph.  If $c_{j} = 1$, the distance between facilities $u_{j-1}$ and
       $u_{j+1}$ becomes $d_{j-1} + d_{j}$ and number of vertices
       between these two facilities can be decomposed as $2(a_{j-1} +
       a_{j}) + 1$ if $b_{j-1} = b_{j} = 0$ or as $2(a_{j-1} + a_{j} + 1) +
       (b_{j-1} + b_{j} -1)$ otherwise. Suppose that this number of
       vertices is decomposed as the latter. We have that, the old payoff of
       player $j$ is $\frac{b_{j-1}}{c_{j-1}+1} + (1 + a_{j-1} + a_{j}) +
       \frac{b_{j}}{c_{j+1}+1}$ and his new one is
       $\frac{b_{j-1} + b_{j} -1}{c_{j-1} + c_{j+1} + 1} +
       \frac{1 + (a_{j-1} + a_{j} + 1) + a_{j+1}}{c_{j+1} + 1} +
       \frac{b_{j}}{c_{j+1} + 1 + c_{j+1}}$. By definition, the
       old payoff is at least $\gamma$ so $a_{j-1} + a_{j} \geq
       \gamma - 1 - \frac{b_{j-1}}{c_{j-1}+1} -
       \frac{b_{j}}{c_{j+1}+1}$. Additionally, by (ii), $a_{j+1} \leq
       \gamma -1$. Thus the old payoff is greater than the new one. Similarly, in
       case the number of vertices between facilities $u_{j-1}$ and $u_{j}$ is decomposed
       as $2(a_{j-1} + a_{j}) + 1$, the old payoff of $j$ is also greater than
       his new one. Hence,
       a player in facility $u_{j}$ has no incentive to move to a
       facility-neighbor $u_{j+1}$.
\end{enumerate}
\end{proof}

\end{document}